\documentclass[%
 aip,
 amsmath,amssymb,
 reprint,%
]{revtex4-1}

\usepackage{pdfpages}
\makeatletter
\AtBeginDocument{\let\LS@rot\@undefined}
\makeatother

\usepackage{tabularx}
\usepackage{dcolumn}
\usepackage{xcolor}
\usepackage{graphicx}
\usepackage{geometry}
\usepackage{gensymb}
\usepackage{cleveref}
\usepackage{bm}
\usepackage[version=4]{mhchem}
\usepackage{multirow}
\usepackage[utf8]{inputenc}
\usepackage[T1]{fontenc}
\usepackage{mathptmx}
\usepackage{footnote}
\usepackage{url}

\graphicspath{ {./fig/} }

\usepackage{geometry}
\begin{document}

\preprint{AIP/123-QED}

\title{MeltNet: Predicting alloy melting temperature by machine learning}

\author{Pin-Wen Guan}
\email{pinweng@andrew.cmu.edu}
\affiliation{Department of Mechanical Engineering, Carnegie Mellon University, Pittsburgh, Pennsylvania 15213, USA}
\author{Venkatasubramanian Viswanathan}
\affiliation{Department of Mechanical Engineering, Carnegie Mellon University, Pittsburgh, Pennsylvania 15213, USA}
\affiliation{Wilton E. Scott Institute for Energy Innovation, Carnegie Mellon University, Pittsburgh, Pennsylvania 15213, USA}
\affiliation{Department of Physics, Carnegie Mellon University, Pittsburgh, Pennsylvania 15213, US}

\date{\today}

\begin{abstract}

Thermodynamics is fundamental for understanding and synthesizing multi-component materials, while efficient and accurate prediction of it still remain urgent and challenging. As a demonstration of the “Divide and conquer” strategy decomposing a phase diagram into different learnable features, quantitative prediction of melting temperature of binary alloys is made by constructing the machine learning (ML) model “MeltNet” in the present work. The influences of model hyperparameters on the prediction accuracy is systematically studied, and the optimal hyperparameters are obtained by Bayesian optimization. A comprehensive error analysis is made on various aspects including training duration, chemistry and input features. It is found that except a few discrepancies mainly caused by less satisfactory treatment of metalloid/semimetal elements and large melting point difference with poor liquid mixing ability between constituent elements, MeltNet achieves overall success in prediction, especially capturing subtle composition-dependent features in the unseen chemical systems for the first time. The reliability, robustness and accuracy of MeltNet is further largely boosted by introducing the ensemble method with uncertainty quantification. Based on the state-of-the-art underlying techniques, MeltNet achieves a prediction mean average error (MAE) as low as about 120 K, at a minimal computational cost. We believe the present work has a general value for significant acceleration of predicting thermodynamics of complicated multi-component systems.
\end{abstract}

\maketitle

\section{Introduction}
Thermodynamics is one of the foundations for understanding materials. It offers the knowledge about phases or structures formed under different condition variables such as composition, temperature, pressure and electric potential, from which people can optimize the conditions to obtain the structures with desirable properties. Thus, predicting thermodynamics of materials has great value in both scientific and practical aspects, although it is also a challenging task due to many variables and underlying mechanisms (lattice disorder, atomic vibration, electronic excitation, magnetic excitation, etc.) involved. So far, there have been two major approaches for this task, ab initio thermodynamic calculations and empirical/semi-empirical calculations represented by the CALPHAD (CALculation of PHAse Diagrams) method. In spite of gaining considerable success, there are still some issues for these approaches. Ab initio thermodynamic calculations are generally time-consuming and vulnerable to errors for complex multi-component systems, and even for a simple binary metallic system, it may lead to incorrect results due to accuracy limitations \cite{Decolvenaere2015}. For the CALPHAD method, one of the disadvantages is that it is highly relied on critical assessment of a blend of data from different sources accomplished by human beings \cite{Lukas2007}, which is not efficient and may be subject to lack of data, and therefore quite challenging for high-throughput modeling of high dimensional systems such as high entropy alloys (HEAs) \cite{miracle2019high}. Therefore, it is urgent to develop a novel approach to model thermodynamics of materials with both efficiency and accuracy. A promising tool to meet this need is machine learning (ML), which has recently shown increasing potential to revolutionize physical sciences including materials science \cite{schmidt2019recent}. Especially, it sheds a light in solving high dimensional problems, which are otherwise almost intractable.

There have been considerable efforts in applying ML to thermodynamics of materials, which can be classified into two categories based on the quantity learned by ML. In the first category thermochemical quantities such as formation enthalpy \cite{schmidt2017predicting, ye2018deep, jha2018elemnet,ubaru2017formation}, Gibbs energy \cite{teichert2019machine} and formation entropy \cite{lapointe2020machine} are learned, based on which the phase diagrams may be obtained \cite{jha2018elemnet}. This category can be also generalized to cover the cases where a customized quantity measuring phase stability is learned, such as site likelihood \cite{ryan2018crystal} and entropy-forming ability \cite{kaufmann2020discovery}. In the second category, the phase equilibrium is learned directly \cite{zhang2020phase,Huang2019, pilania2015structure, seko2014machine}. The first category is obviously more physics-based, but thermochemical quantities are usually complicated and may be challenging to learn for multi-component non-stoichiometric phases. In addition, in both categories, temperature is usually absent in the models in practice, which is an apparent drawback from the perspective of thermodynamics.  

Predicting phase diagrams of complicated multi-component systems directly by ML is also challenging. To reduce the difficulty, the old wisdom “Divide and conquer” may be a viable strategy, i.e., decomposing a phase diagram into different features learned by different ML models respectively. As a first attempt to realize this strategy, quantitative prediction of melting temperature of binary alloys is made by constructing a ML model termed “MeltNet” in the present work. Melting temperature is an important phase equilibrium feature defining the boundary between two fundamental states of condense matter, as well as a critical parameter in many important applications such as solder materials, metallic glass \cite{dasgupta2019probabilistic} and room-temperature liquid metal electrodes (LME) for rechargeable batteries \cite{guo2019room}. It is noted that there have been a few works on predicting melting points of binary compounds by ML \cite{pilania2015structure, seko2014machine}, but they have very strict constraints on the stoichiometry, and therefore lacks generalizability in arbitrary chemical compositions present in the MeltNet.

\section{Computational methods}
\subsection{Data generation}
\begin{figure}
\includegraphics[width=\linewidth]{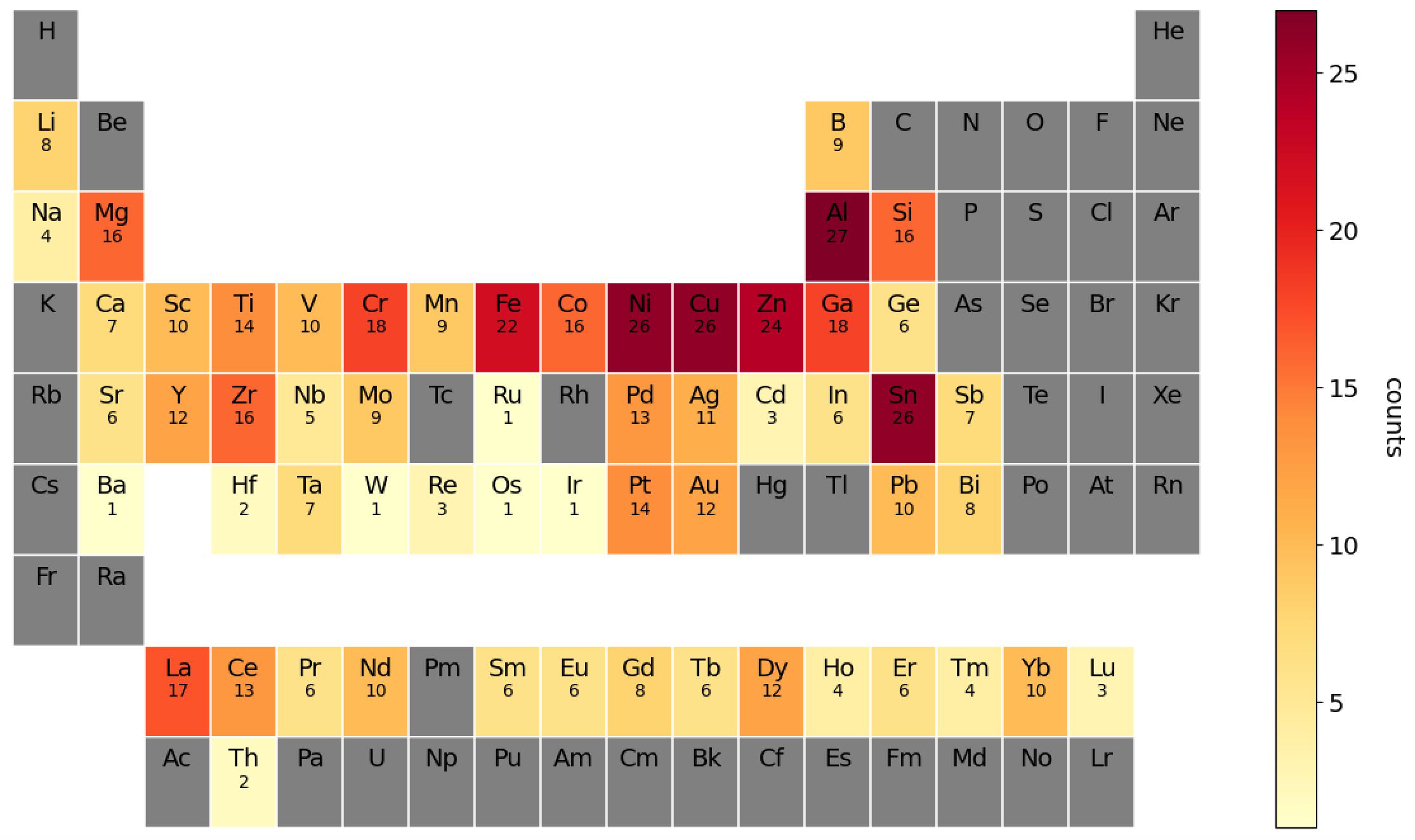}
\centering
\caption{Distribution of elements in the systems studied in the present work. The number under the name of each element represents the times of that element appearing in the studied systems, which is also indicated by the color. The elements with grey color are not involved in the present work.} 
\label{fig:ele}
\end{figure}

For alloy systems, the melting temperature is generally an interval bounded by the liquidus and the solidus instead of a single value. To simplify the problem, the treatment by Chelikowsky and Anderson \cite{Chelikowsky1987} is adopted in the present study, i.e., defining the liquidus as the melting temperature. The tdb files are retrieved from NIMS Computational Phase Diagram Database (https://cpddb.nims.go.jp/en/), from which the liquidus of each binary system is calculated using pycalphad \cite{Otis2017}. There are 287 different binary alloy systems collected in total, involving 57 elements with distribution shown in \cref{fig:ele}. The non-metal elements are absent except B and Si, but the methodology described in the present work should be also applicable to them. The composition is sampled in a step of 0.01 from 0 to 1 with the endmembers excluded, resulting 99 samples per system. However, a tiny portion of calculations are abnormal and therefore removed. Finally, a total of 28148 data was generated, which consist of a large database for the subsequent study.

\subsection{ML model}

The deep neural network (DNN) is employed as the ML model and termed MeltNet in the present work, and is implemented in PyTorch \cite{Paszke2017}. Seven descriptors are used as the inputs for MeltNet, including the valence electron concentration VEC, electronegativity difference $\Delta\chi$, atomic radius difference $\delta$, ideal mixing entropy $\Delta S_{mix}$, formation enthalpy $\Delta H_{f}$, average fusion entropy $S^{fus}$, fusion entropy weighted average melting point $\tilde{T}$. Among them, VEC, $\Delta\chi$, $\delta$ and $\Delta S_{mix}$ adopt the definitions in the reference \cite{Huang2019}, while $\Delta H_{f}$ is taken from the Materials Project database \cite{Jain2013}. The following definitions are used for a n-component system:

\[S^{fus}=\sum_{i=1}^{n}c_{i}S^{fus}_{i}\]

\[\tilde{T}=\frac{\sum_{i=1}^{n}c_{i}S^{fus}_{i}T_{i}}{S^{fus}}\]

\noindent where $c_{i}$, $S^{fus}_{i}$ and $T_{i}$ are the concentration, fusion entropy and melting point of the element i, respectively. Some rationales can be given for selecting the above descriptors. VEC, $\Delta\chi$, $\delta$ and $\Delta H_{f}$ are all relevant to stability of the solid phases. $\Delta S_{mix}$ is related to stability of both solid and liquid solutions. $S^{fus}$ describes the basic part of fusion entropy contributed by the linear mixing between elements. $\tilde{T}$ would be the true melting temperature if both fusion entropy and fusion enthalpy contain only the contribution from the linear mixing between elements. It is noted that the definitions of all the descriptors are general for any multi-component system without restriction on the number of components, although only binary systems are studied in this work.   
In addition, for convenience in later discussions, the average melting point based on the Vegard’s law is defined by

\[\bar{T}=\sum_{i=1}^{n}c_{i}T_{i}\]

\noindent and the excess melting point, i.e., the deviation of the true melting point $T$ from $\bar{T}$ is defined by

\[\Delta T=T-\bar{T}\]

\noindent which is the object variable to be learned by MeltNet. The error or the loss function is defined as the L1-norm loss, i.e.,

\[\epsilon=\sum_{i=1}^{m}\frac{|\widehat{\Delta T_{i}}-\Delta T_{i}|}{m}\]

\noindent where m is the number of data in the dataset, and $\widehat{\Delta T_{i}}$ is an estimation for $\Delta T_{i}$.

\subsection{Bayesian optimization of hyperparameters}
\begin{figure*}
\includegraphics[width=\textwidth]{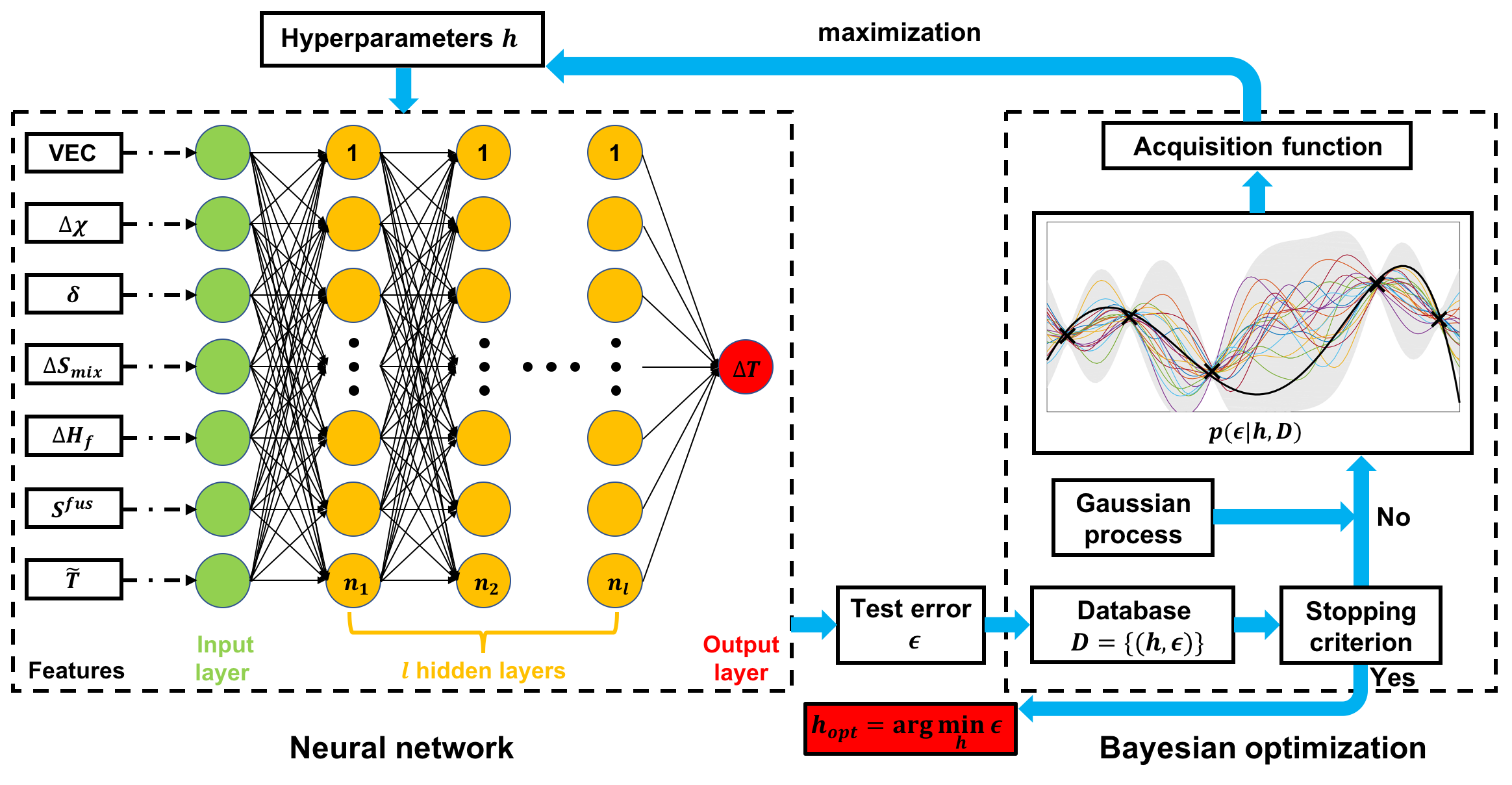}
\centering
\caption{Architecture of MeltNet with Bayesian optimization of hyperparameters. The input features are described in the main text.} 
\label{fig:nn}
\end{figure*}

There are multiple hyperparameters in MeltNet and its training process, with the important ones including: (1) number of hidden layers, (2) number of nodes in each layer, (3) momentum, (4) batch size, (5) learning rate, and 6) weight of decay.
These hyperparameters form a high-dimension space, for which the optimization is quite challenging. To accomplish this task, a very useful black-box global optimization technique, the Bayesian optimization method was employed by using the Dragonfly package \cite{Kandasamy2020}. In the Bayesian optimization, every evaluation of the objective function (the test error as a function of hyperparameters here) at some point is used to update the posterior distribution over the objective function, from which an acquisition function is constructed to determine the next point of evaluation. The posterior distribution over the objective function is calculated based on the Gaussian process. The optimization is implemented in two stages. In stage 1, the six hyperparameters listed above are optimized simultaneously, assuming the number of nodes in each layer is a constant. In stage 2, all the hyperparameters are fixed at the optimized values obtained from stage 1, except that the number of nodes in each layer can be changed independently and treated as optimization variables. The architecture of MeltNet with Bayesian optimization of hyperparameters used in the present work is illustrated in \cref{fig:nn}.

\subsection{Uncertainty quantification}
The ML model can be largely influenced by the choice of samples for training. To quantify the uncertainty associated with such choice, the ensemble approach was employed here, which has been applied successfully in other fields of computational materials science, e.g., the Bayesian Error Estimation Functional (BEEF) where an ensemble of density functionals are used to estimate the exchange-correlation errors \cite{PhysRevB.85.235149}. Especially, the ensemble approach has been applied in thermochemical properties \cite{Guan2019} and phase diagrams \cite{yuan2020uncertainty,guan2020mathcalp2}. In the present work, the training dataset was sampled randomly for 100 times forming 100 subsets, and each subset contains 75\% of the total training data. Each subset was used for training a set of MeltNet parameters. As a result, an ensemble of 100 sets of MeltNet parameters can be obtained, providing an ensemble of 100 predictions. The average and standard deviation of the prediction ensemble can then be calculated, with the latter taken as the uncertainty of the prediction. 

\section{Results and discussion}
\subsection{Baseline prediction and trends in melting temperatures}
\begin{figure}
\includegraphics[width=\linewidth]{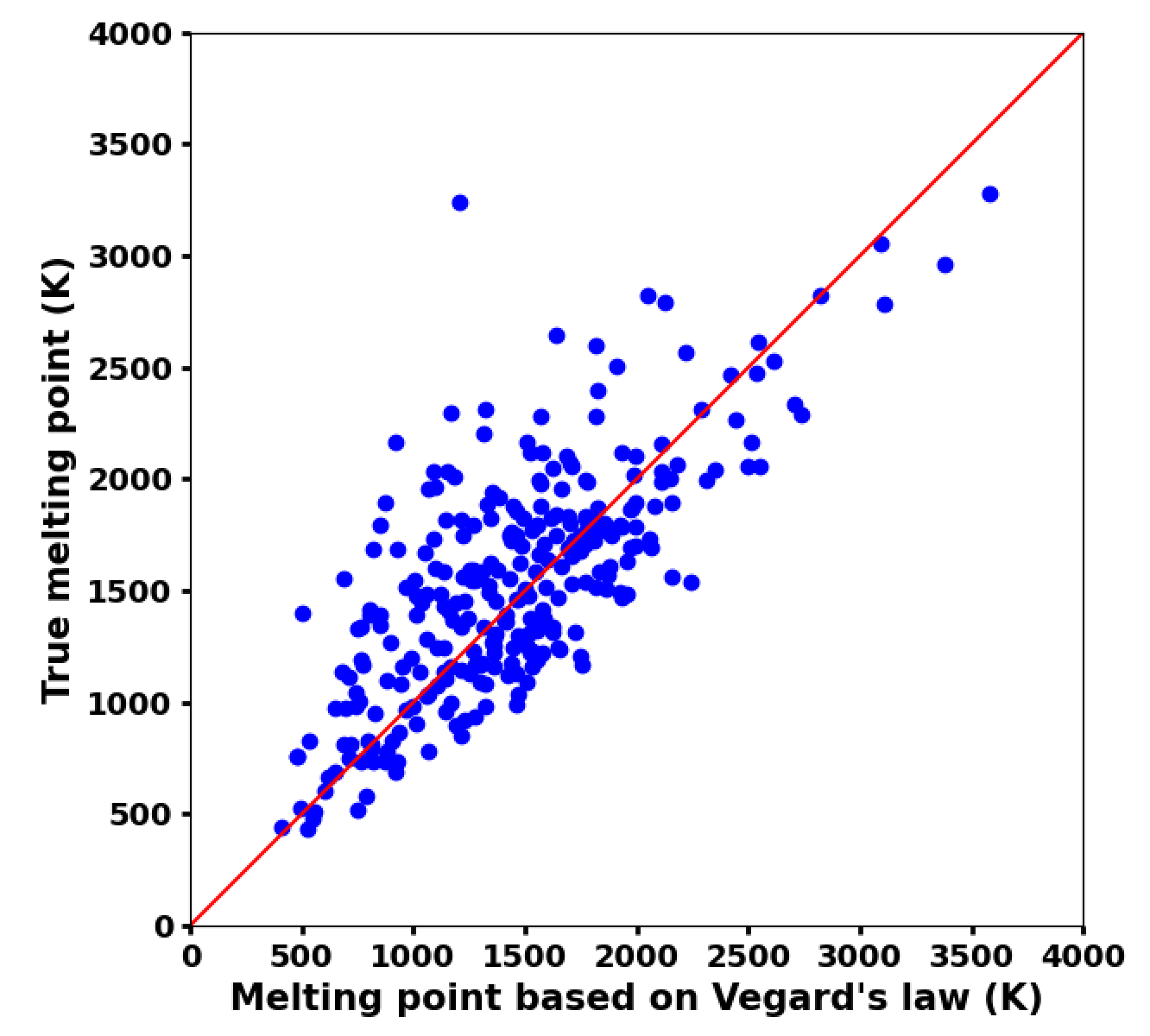}
\centering
\caption{Parity plot between the true melting point and the prediction based on Vegard’s law. The shown data points are those with equal molar composition.} 
\label{fig:vegard}
\end{figure}

The simplest prediction one can imagine is probably the average melting point based on the Vegard’s law $\bar{T}$, which is considered as the baseline in this work. \cref{fig:vegard} shows the relationship between $\bar{T}$ and the true melting point $T$. Due to the large amount of the total data, only the data with equal molar composition is shown. The MAE of the Vegard’s prediction on the whole dataset is 218 K, which is quite significant. It is noted that the deviation of  $\bar{T}$ from $T$ is not symmetric, and $\bar{T}$ has a larger tendency to be an underestimation of $T$ due to the physical constraint that $T$ is always above 0 K while it has no apparent upper limit.

\begin{figure*}
\includegraphics[width=\linewidth]{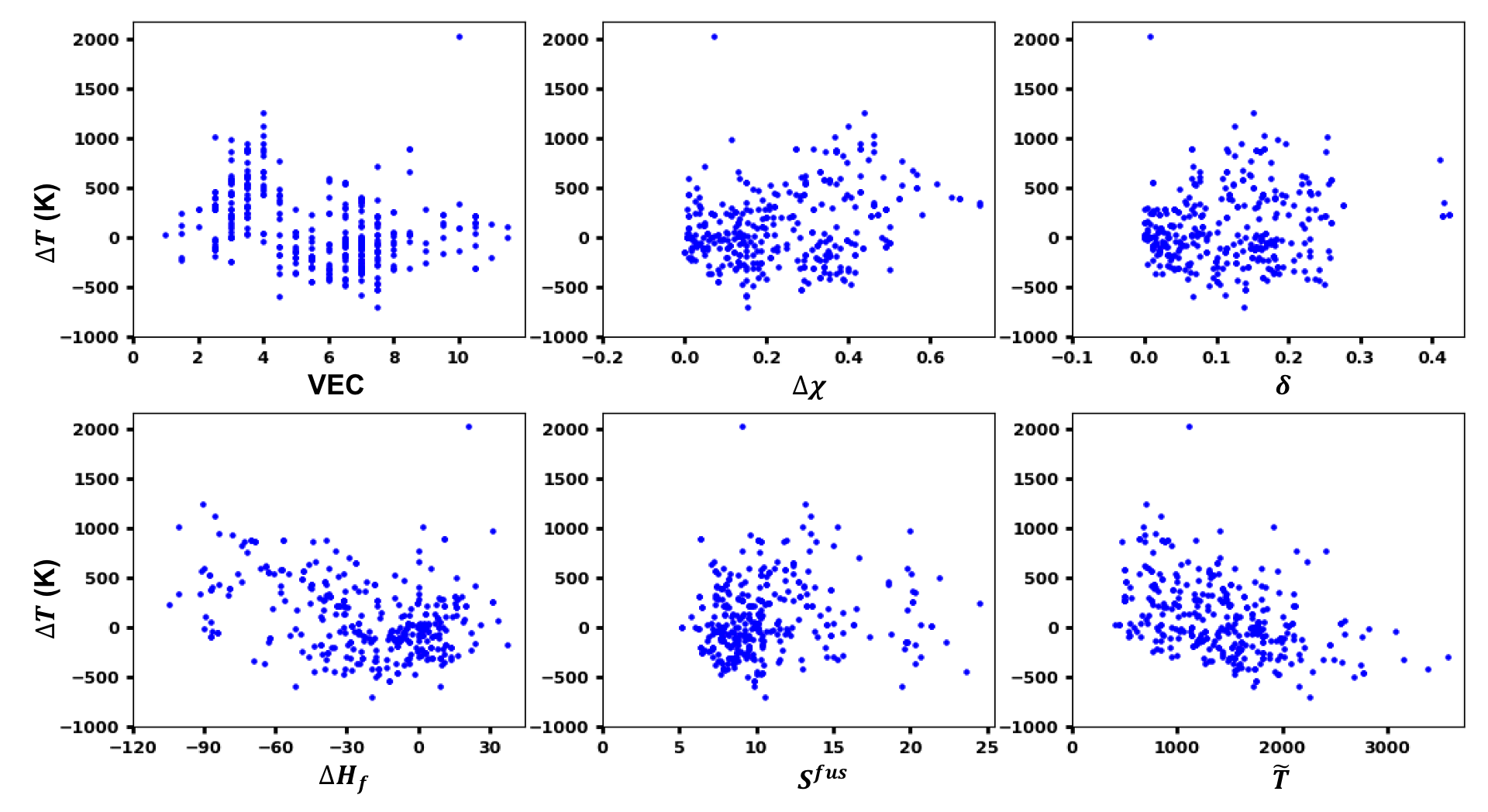}
\centering
\caption{Relation between the excess melting temperature and the input features. The shown data points are those with equal molar composition.} 
\label{fig:dt}
\end{figure*}

Before the ML prediction, it is beneficial to have a rough overview about the relation between the melting point and each individual feature (\cref{fig:dt}). Due to the large amount of the total data, only the data with equal molar composition is shown, and the mixing entropy $\Delta S_{mix}$ is therefore absent from the shown features. It is found that the excess melting point $\Delta T$ has weak one-to-one correlation with $\Delta\chi$, $\delta$ and $S^{fus}$, which is somewhat against intuition, since at least $\Delta\chi$ and $\delta$ are considered as important factors affecting stability of compounds and alloys and therefore the melting point. On the other hand, lack of one-to-one correlation does not necessarily mean lack of correlation, which may be via complicated synergy with other features. For the three features, VEC, $\Delta H_{f}$ and $\tilde{T}$, negative one-to-one correlation with $\Delta T$ can be observed, though not highly significant. An intuitive explanation is that, high-VEC elements, e.g., transition metals, usually have high melting points, making the baseline $\bar{T}$ high, therefore the “chance” to have more negative $\Delta T$ is higher. The effect of $\tilde{T}$ can be also understood in a similar way. The effect of $\Delta H_{f}$ is more obvious: a more negative $\Delta H_{f}$ means stronger stability of the solid at the corresponding composition, and therefore more positive $\Delta T$. A previous work attempted to correlate binary alloy melting points with chemical coordinates consisting of 10 properties such as atomic radii and bulk moduli (VEC was not included), but only poor correlation was found except for cohesive energies and elemental melting points \cite{Chelikowsky1987}, which is in consistence with the present observations.

\begin{figure*}
\includegraphics[width=\linewidth]{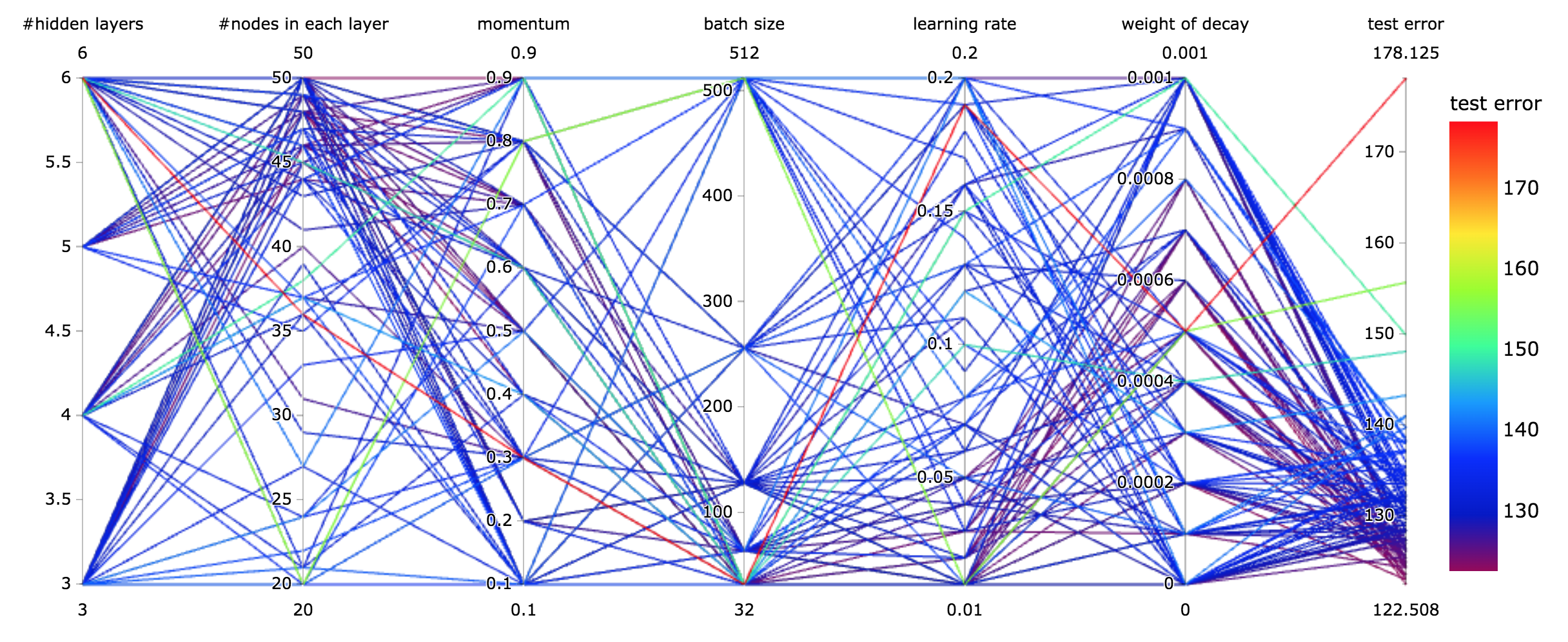}
\centering
\caption{Relation between the test errors (in K) and the hyperparameters of MeltNet.} 
\label{fig:hyper}
\end{figure*}

\begin{figure*}
\includegraphics[width=\linewidth]{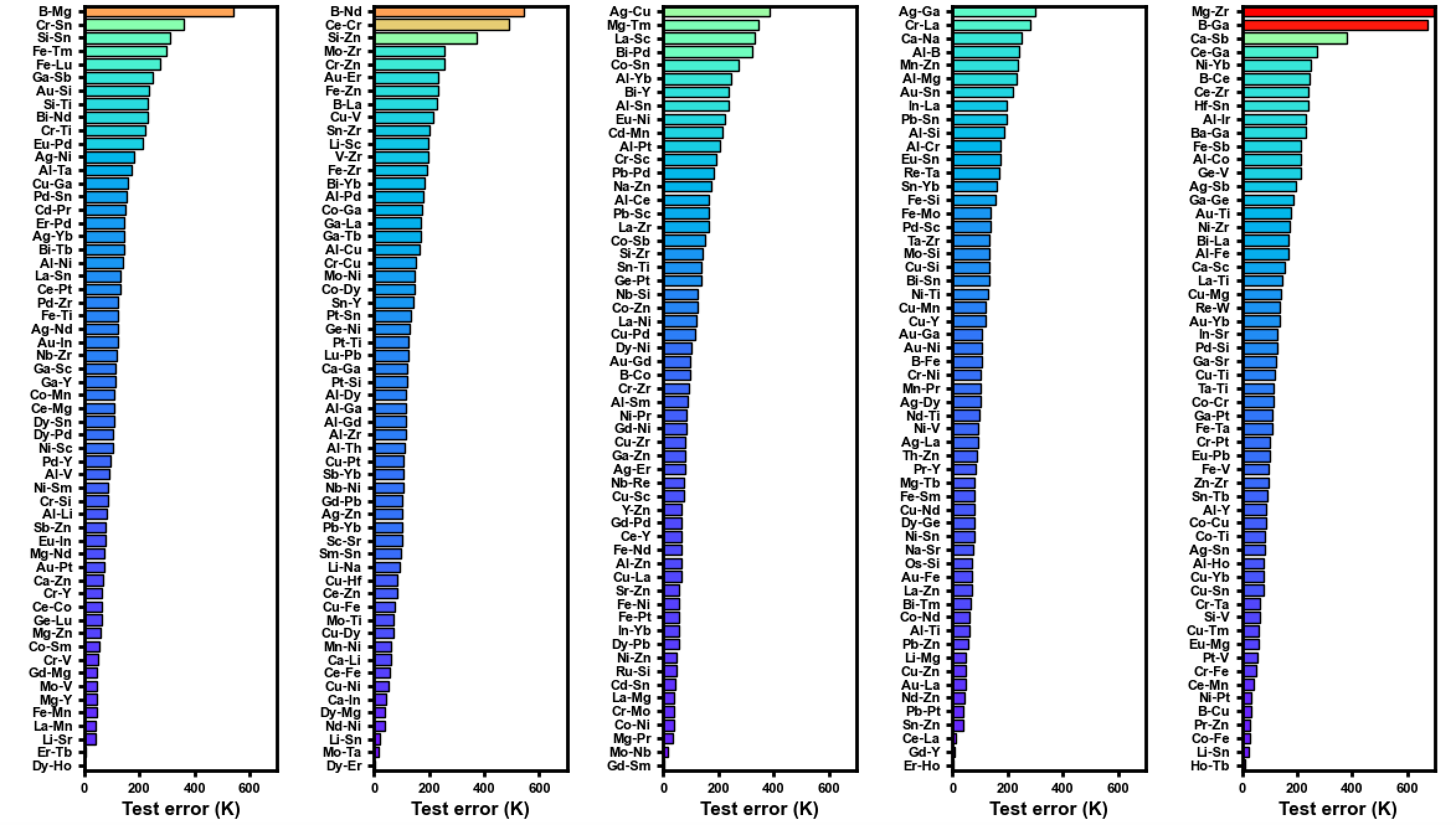}
\centering
\caption{Test errors for each system in the five training-test dataset pairs, in each of which one subset was used as the test set and the other four subsets formed the training set together.} 
\label{fig:error-sys}
\end{figure*}

\begin{figure}
\includegraphics[width=\linewidth]{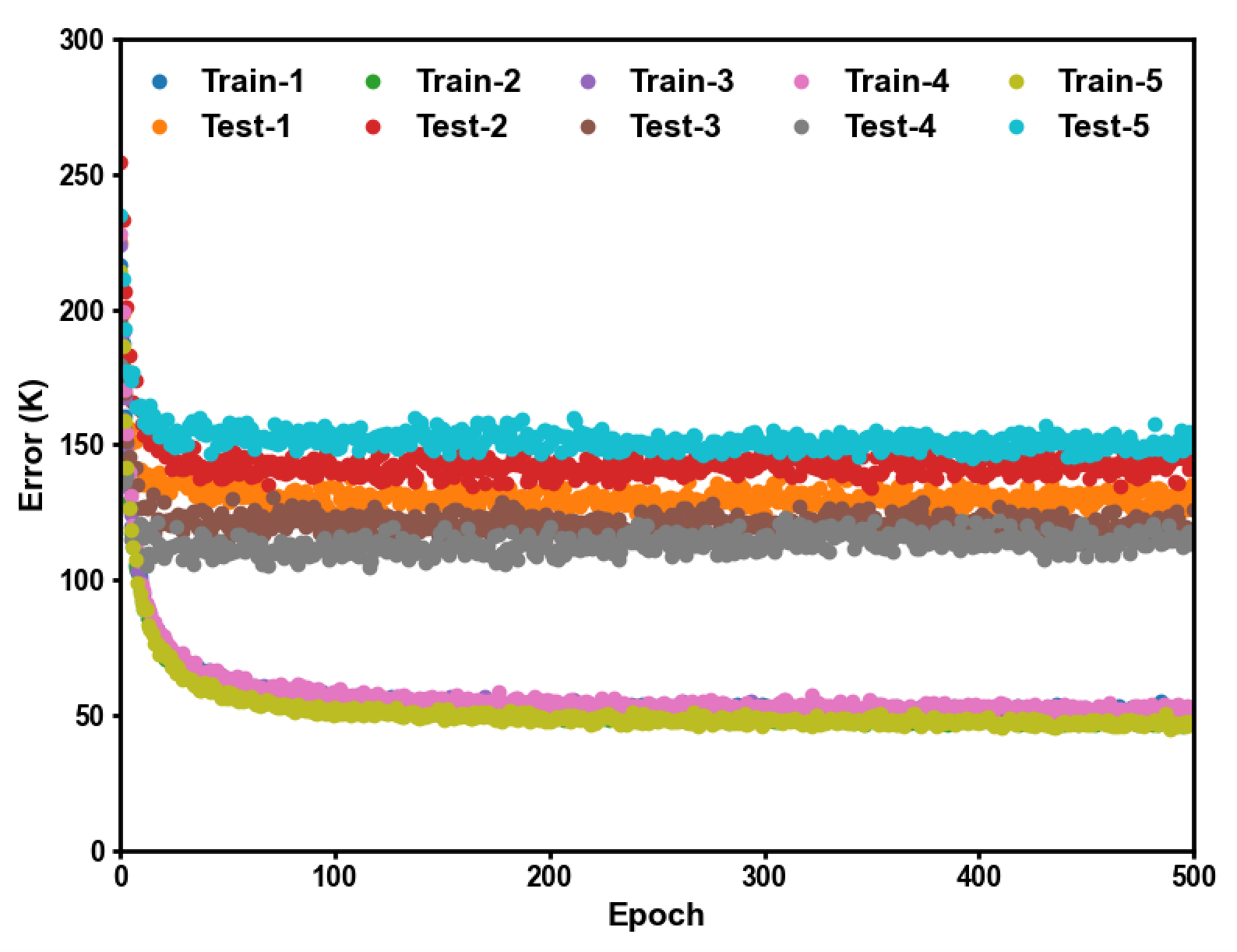}
\centering
\caption{Evolution of errors during training for the five training-test dataset pairs.} 
\label{fig:error-epoch}
\end{figure}

\begin{figure*}
\includegraphics[width=\linewidth]{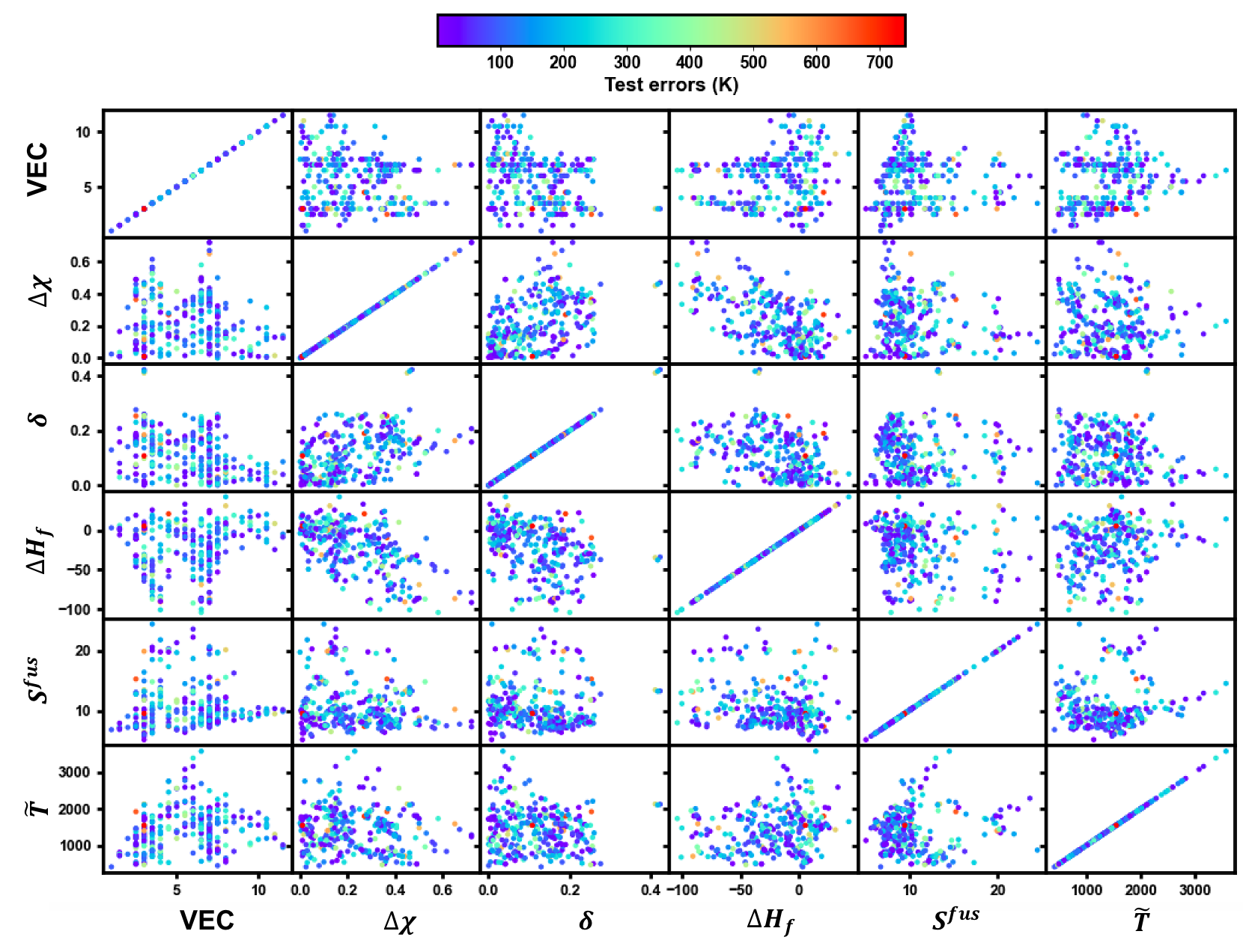}
\centering
\caption{Relation between the test errors and the combination of any two input features. The shown data points are those with equal molar composition.} 
\label{fig:error-feature}
\end{figure*}

\begin{figure*}
\includegraphics[width=\linewidth]{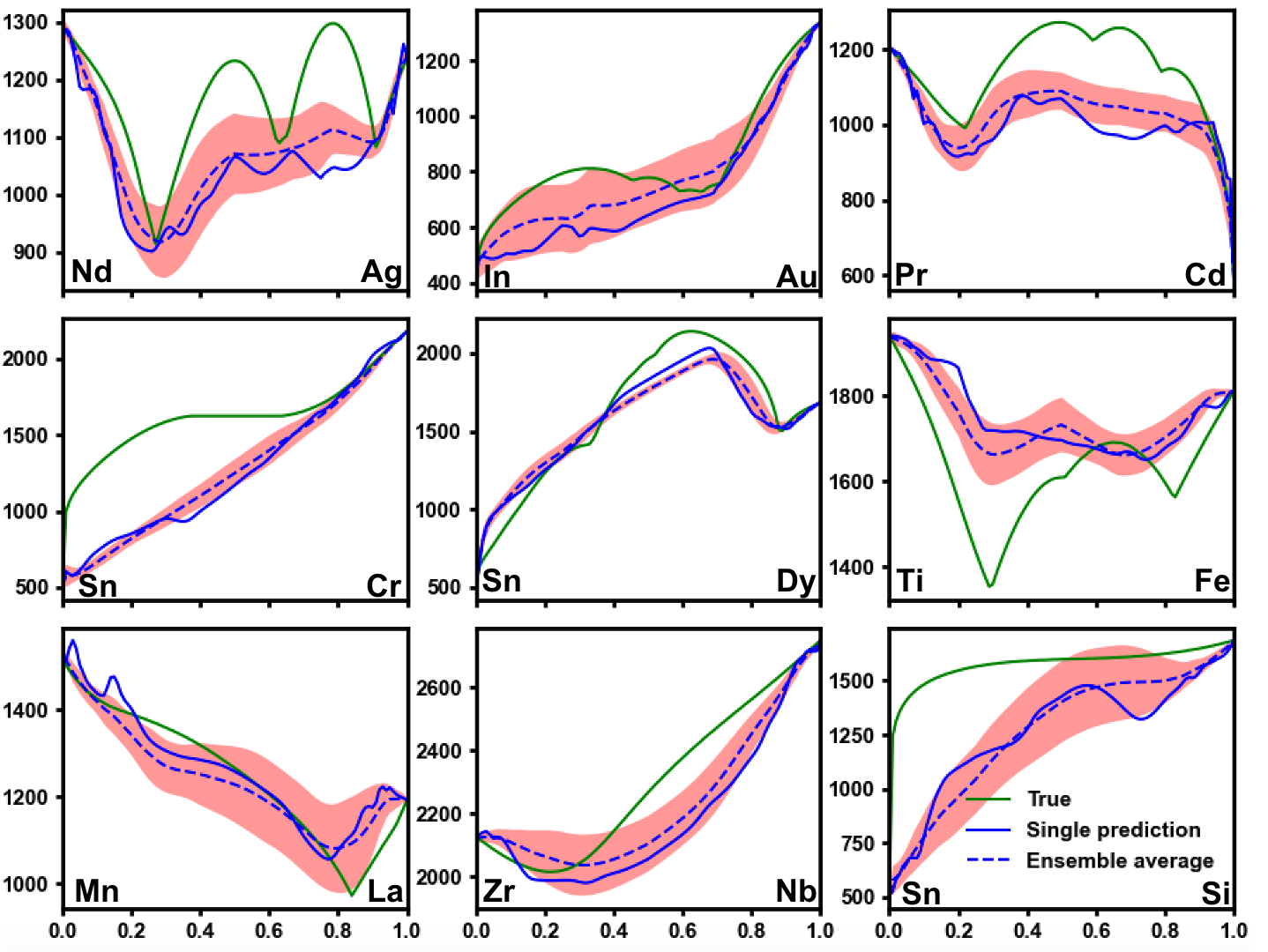}
\centering
\caption{Predicted vs true melting point for nine sample systems. The red shadow area represents uncertainty associated with the choice of training data in the ensemble prediction.} 
\label{fig:melt}
\end{figure*}

\subsection{Optimization, training and factors affecting errors}
In stage 1 of Bayesian optimization of hyperparameters of the DNN model where the number of nodes in each hidden layer is equal, the optimization was run for 100 iterations. The searching space is 20-50 for the number of nodes per hidden layer, 3-6 for the number of hidden layers, 0.1-0.9 for the momentum, $2^n$ (n=5,6,7,8,9) for the mini-batch size, 0.01-0.2 for the learning rate, and $10^{-4}-10^{-3}$ for the weight-decay factor. Based on the data generated in the optimization, the relation between the test error and the hyperparameters is visualized in the parallel coordinates diagram (\cref{fig:hyper}). It should be noted that the space of hyperparameters is not homogeneously sampled. Instead, the space region with higher probability to minimize the test error is more likely sampled due to the algorithm of Bayesian optimization \cite{Kandasamy2020}. Interestingly, though the depth of a DNN is generally considered as one of its important advantages, it is found that the test error is not sensitive to the number of hidden layers in the present case. In contrast, the number of nodes per hidden layer has very significant influences on the test error. To minimize the test error, a relatively large number of nodes per hidden layer is preferred. A small enough mini-batch size and a low learning rate are also preferred, while the momentum and the weight-decay factor does not matter too much. Finally, a 3-hidden-layer DNN with 48 nodes per hidden layer trained with 32 of the mini-batch size, 0.01 of the learning rate, 0.5 of the momentum and $6*10^{-4}$ of the weight-decay factor is found to be optimal in the search. In stage 2, the numbers of nodes of different hidden layers were allowed different, but no further decrease of the test error was detected. Therefore, the optimal DNN from stage 1 is adopted as the working model of MeltNet in the present study.

To avoid bias introduced by the partition of the training/test sets, the cross validation method was employed, where the total 287 binary systems were randomly partitioned into five subsets, leading to five training-test dataset pairs, in each of which one subset was used as the test set and the other four subsets formed the training set together. \cref{fig:error-epoch} shows the evolution of training and test errors in the training process for the five training-test dataset pairs. It can be seen that the test errors have modest variation between the dataset pairs, but the training errors are very similar. After rapid decrease in the initial tens of epochs, the training errors continue to decrease slowly, but the test errors almost stop decreasing, and some dataset pairs even exhibit overfitting behavior. 

To further clarify how the test errors depend on the underlying chemistry, the system-resolved test MAE is calculated for all the 287 systems and ranked within the test set of the corresponding dataset pairs (\cref{fig:error-sys}). The MeltNet predictions for the majority of the studied systems have a test MAE within about 200 K, indicating an improvement on the baseline average MAE, 218 K, which demonstrates the success of MeltNet in predicting the unseen chemistry. However, the MeltNet predictions are still not satisfactory for some systems, with five systems B-Mg, B-Nd, Ce-Cr, Mg-Zr and B-Ga being the most problematic. Interestingly, three out of these five systems contain the element boron, a metalloid with properties between those of metals and non-metals. It is worth noting that for the B-Nd system, there are considerable discrepancies between different sources in the literature. The true value adopted in the present study is calculated based on the work by Hallemans et al. \cite{Hallemans1995}, where the liquidus line in the intermediate and Nd-rich compositions is higher than that in a work \cite{Liao1996} published later by several hundreds of kelvin. In other words, more experimental and theoretical efforts are needed to clarify the origin of the large prediction error for the B-Nd system. The Mg-Zr system and the Ce-Cr system both consist of a low-melting-point element and a high-melting-point element, as well as have weak liquid mixing ability characterized by the liquid miscibility gap. The present model fails to capture such scenario, which will be subject to future study and improvement. 
  
For better understanding of the origins of the errors, the data with equal molar composition is plotted in the pairplot of the features, with the test error represented by the color of the data point. There is no significant correlation between the test error and the features, except that it is quite evident that a higher fusion entropy averaged from the constituent elements by their compositions, $S^{fus}$, tends to lead to larger test errors. According to Fig. S1 (see Supplementary material), the elements with high fusion entropy can be roughly classified into two large classes, metalloids/semimetals (e.g., Bi, Sn, Ge, Si, Ga, B, Sb) and refractory metals (e.g., Os, Ru, Ir, Mo, W). The cases where large test errors occur usually belong to the first class. Obviously, the chemical nature of the systems containing metalloids/semimetals are quite different from that of the systems containing metals exclusively, making the predictions more challenging. 

\subsection{Predictions by MeltNet}
The predicted melting points of selected systems are presented in \cref{fig:melt} as examples, with the true melting points adopted in the present work also shown. These selected systems are sampled from the test set of the first training-test dataset pair ranked in alphabetical order with the equal sampling interval to reduce possible bias. In the single prediction, the whole training set was used to train a single DNN model resulting in a single prediction, while in the ensemble prediction, the whole training set was sampled to generate multiple subsets to train an ensemble of DNN models resulting in an ensemble of predictions. The ensemble standard deviation is used to quantify the uncertainty associated with the training data, which is represented by the red shadow area. It can be seen that for most systems, both types of predictions achieve general agreement with the true values, at least semi-quantitatively. Especially, the kinks in the melting point curve are usually well captured by the predictions, which is quite non-trivial and encouraging, provided these test systems are unseen for MeltNet. To the best of our knowledge, successful ML prediction of such subtle composition-dependent features of thermodynamic equilibrium is unprecedented. It is also worth noting that although the present dataset is quite large, the number (~230) of systems used for training is still a small portion (~14\%) compared with the number (1596) of all the binary systems that can be formed from the studied 57 elements, implying good generalizability of MeltNet. Admittedly, some quantitative discrepancies still exist. For example, in the Ag-Nd system, the “valley” in the Nd-rich side and the “twin peaks” in the intermediate compositions and Ag-rich side are all captured by the predictions and the depth of the “valley” is also well predicted, but the height of the “twin peaks” is underestimated. In addition, the predicted positions of the kinks sometimes have minor shift compared with the true values. The most unsatisfactory performance of the present model is for the Cr-Sn and Si-Sn systems, both of which are combinations of a low-melting-point element and a high-melting-point element with weak liquid mixing ability and therefore cannot be well reproduced by the present predictions as discussed above. 

In terms of the two types of predictions, it is found that the ensemble prediction is obviously superior to the single prediction. The ensemble prediction provides uncertainty guiding the decision that how much confidence should be given to the prediction. For example, in the Au-In system, there are considerable errors in the prediction of the “bump” in the In-rich side, but the large uncertainty reminds one to prudently assess the prediction, while the small uncertainty in the Au-rich side is in line with the high accuracy of the prediction. However, it should be pointed out that the uncertainty here is only associated with the training data, which does not cover all the sources of uncertainty such like the model itself. Thus, it is not guaranteed that the true values can be always bounded by the uncertainty of the present ensemble prediction, as shown by the Cr-Sn and Si-Sn systems. Another significant advantage of the ensemble prediction is that it is more robust and less noisy, as evident by the smooth behavior of the ensemble average compared with the spurious behavior of the single prediction. In fact, as shown in Table 1, the ensemble average has considerably smaller test MAE (122.5 K) than the single prediction (133.1 K) in overall and performs invariably better in each test in the five-fold cross validation. The average uncertainty of the ensemble prediction is 74.7 K in overall and does not change too much across the tests in the cross validation. Based on these observations, it is implied that the ensemble approach should be routinely employed in the ML prediction of thermodynamics. Compared with the test MAE by the baseline prediction based on the Vegard’s law (218.0 K), MeltNet significantly improves the prediction accuracy, at expense of little increase in computational costs, which is vital important for complicated multi-component systems.

\section{Conclusions}
To make complicate phase equilibria of multi-component systems learnable, the “Divide and conquer” strategy is proposed in the present work, where the whole phase diagram is decomposed into different phase equilibria features which are relatively easy to learn by ML. As one of the most important phase equilibria features, the melting point is chosen as the first example to demonstrate the state-of-the-art methodology for such kind of problems. MeltNet, a DNN model with seven input features is constructed, with the optimal hyperparameters obtained from Bayesian optimization. The number of nodes per hidden layer, mini-batch size and learning rate are found crucial for the model performance. Five-fold cross validation is employed to reduce bias related to training-test dataset partition. A thorough analysis is made for the dependence of the prediction errors on various aspects including hyperparameters, training duration, chemistry and input features. It is found that large prediction errors mainly originate from less satisfactory treatment of metalloid/semimetal elements and large melting point difference with poor liquid mixing ability between constituent elements. Despite a minority of failures and some quantitative discrepancies, MeltNet offers satisfactory predictions in general, especially capable of capturing subtle composition-dependent features successfully for the first time. It is also found that the ensemble prediction is more reliable, robust and accurate with uncertainty quantification. The prediction MAE by MeltNet is as low as about 120 K, at a negligible computational cost compared with other methods. We believe the present work sets a solid cornerstone for ML of thermodynamics of complicated multi-component systems.

\begin{table*}
\begin{tabularx}{0.8\textwidth}{ >{\centering\arraybackslash}X>{\centering\arraybackslash}X>{\centering\arraybackslash}X>{\centering\arraybackslash}X>{\centering\arraybackslash}X>{\centering\arraybackslash}X>{\centering\arraybackslash}X } 
\hline
 & Test 1 & Test 2 & Test 3 & Test 4  &  Test 5 &  Overall \\
\hline
Size & 5683 &	5672 &	5599 &	5604 &	5590	& 28148  \\
$\epsilon_{baseline}$ (K) & 225.2 &	254.5 &	195.5 &	179.5 &	235.0 &	218.0 \\
$\epsilon_{MeltNet}^{single}$ (K) & 130.8 &	143.3 &	125.8 &	116.6 &	148.9 &	133.1 \\
$\epsilon_{MeltNet}^{ensemble}$ (K) & 123.8 & 133.3	& 109.6 &	104.2 &	141.7 &	122.5 \\
$\sigma_{MeltNet}^{ensemble}$ (K) & 70.1 & 75.5 &	75.7 & 77.8	& 74.4 & 74.7 \\
\hline
\end{tabularx}
\caption{\label{tab:compare}Comparison between different methods in predicting melting temperatures for the five training-test dataset pairs, listing test errors by baseline prediction based on the Vegard’s law $\epsilon_{baseline}$, single prediction by MeltNet $\epsilon_{MeltNet}^{single}$, and ensemble average prediction by MeltNet $\epsilon_{MeltNet}^{ensemble}$. The size of each test dataset and the uncertainty associated with the choice of training data in the ensemble prediction $\sigma_{MeltNet}^{ensemble}$ are also listed.}
\end{table*}

\section{Supplementary Material}

See Supplementary Material for detailed descriptions of data used in this work, and the results not covered in the main text. 

\section{Acknowledgement}

This work was partially supported by the Assistant Secretary for Energy Efficiency and Renewable Energy, Office of Vehicle Technologies of the U.S. Department of Energy (DOE)
through the Advanced Battery Materials Research (BMR) Program under Contract No. DE-EE0007810. Acknowledgment is also made to the Extreme Science and Engineering Discovery Environment (XSEDE) for providing computational resources through Award No. TG-CTS180061.

\clearpage

\bibliography{cite}

\includepdf[pages={{},1,{},2}]{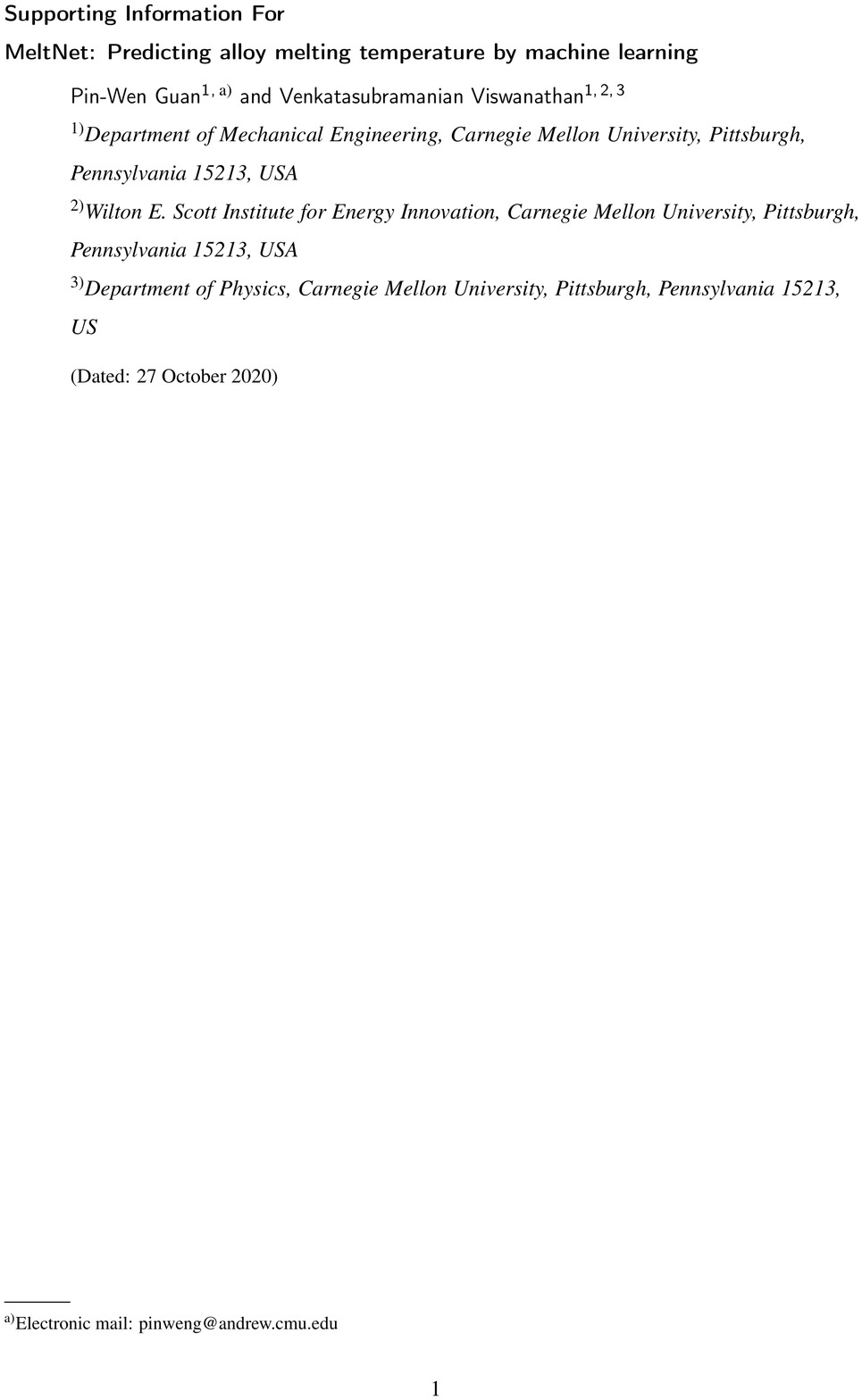}

\end{document}